# Carrier transport and electron-lattice interactions of nonlinear optical crystals $CdGeP_2$, $ZnGeP_2$ and $CdSiP_2$

Rishmali Sooriyagoda,[1] Herath P. Piyathilaka,[1] Kevin T. Zawilski,[2]
Peter G. Schunemann[2] and Alan D. Bristow[1]

[1]Department of Physics and Astronomy, West Virginia University, Morgantown
West Virginia 26501-6315, U.S.A.
[2]BAE Systems, MER15-1813 P.O. Box 868, Nashua, New Hampshire 03061, U.S.A

Terahertz time-domain spectroscopy is employed to investigate temperature-dependent properties of bulk chalcopyrite crystals ($CdGeP_2$, $ZnGeP_2$ and $CdSiP_2$). The complex spectra provide the refraction and absorption as a function of temperature, from which electron-phonon coupling and average phonon energies are extracted and linked to the mechanics of the A- and B-site cations. Also, AC conductivity spectra provide carrier densities and electron scattering times, the temperature dependence of which is associated with unintentional shallow dopants. Temperature dependence of the scattering time are converted into carrier mobility and modeled with microscopic transport mechanisms such as polar optical phonon, acoustic phonons, deformation potential, ionized impurity and dislocation scattering. Hence, analysis links the THz optical and electronic properties to relate microscopic carrier transport and carrier-lattice interactions.

**Introduction**

The traditional contact-based Hall-effect measurement is a powerful tool, commonly used to investigate electrical transport. It provides a wealth of information about materials and their devices, classifying microscopic transport mechanisms and even distinguishing classical or quantum transport behavior. Over the past few decades the enormous improvement in contactless terahertz time-domain spectroscopy (THz-TDS) [1–3] has opened a window of accurate non-invasive optical and transport data for liquids, gases and solids [4–6]. In the latter, THz-TDS is used to provide feedback for material growth and measure device performance. Terahertz (THz) pulses measure carrier dynamics on picoseconds time scale [7] through low-energy intraband excitation of free carrier. One advantage of this method is that the electric field of the THz pulses can be measured directly, such that a Fourier transform from time domain gives a complex response spectrum. Thus, the material or devices can be analyzed through the refractive index, extinction coefficient, dielectric function and AC conductivity without performing Kramers-Kronig analysis [4,8,9]. The complex refractive index can be related to lattice vibrations and the AC conductivity to microscopic scattering mechanism [10,11].

There is still much to understand about models of conduction and their effect at THz frequencies, which requires application of THz-TDS to bulk crystals without contacts or other forms of device processing. One material class that has been interesting to the optics community is II-IV-$V_2$ chalcopyrite crystals due to their high nonlinearity, composition-tunable band gaps, wide transparency windows and high damage threshold [12–15]. Chalcopyrite crystals have been explored for electromagnetic (EM) screening, spintronic [16,17], and photovoltaic applications [18], making them good as optical and optoelectronic materials. Large-area chalcopyrite crystal growth offers the opportunity to produce devices [19–23], especially frequency converters. While optical properties are generally known in these materials, there have been no prior systematic studies of temperature-dependent THz dispersion and AC conductivity. Hence, chalcopyrite crystals are a model system for exploring temperature-dependent THz-TDs acquisition of microscopic electronic transport phenomena by applying models that have traditionally been applied to Hall measurements as a function of temperature. This approach overcomes poor experimental reliable of Hall measurements of chalcopyrite crystals due to their low mobility.

In this paper, temperature-dependent THz-TDS is reported for Cadmium germanium diphosphide ($CdGeP_2$), zinc germanium diphosphide ($ZnGeP_2$) and cadmium silicon diphosphide ($CdSiP_2$). It is found that the refractive index increases with the increasing temperature associated with greater electron-phonon coupling. By fitting complex AC conductivity to the Drude-Smith model, the carrier density and scattering time are obtained. Temperature dependence of the carrier density is described by the Fermi-Dirac model for carrier occupancy and to identify dopants. The carrier mobility is determined from the reciprocal addition of extracted carrier scattering times, giving a temperature-dependent mobility that is described by a combination of polar optical phonon, acoustic phonons, deformation potential, ionized impurity and dislocation scattering.

**Experimental Details**

Undoped single crystals of $CdGeP_2$ (CGP), $ZnGeP_2$ (ZGP), and $CdSiP_2$ (CSP) are grown using the horizontal gradient freeze technique [24]. Samples were cut in the (110) plane into 1 cm × 1cm × 0.5 mm chips and double-side polished for optical measurements. The samples were placed in an optical cryostat so their temperature could be controlled between 4 K and 300 K for THz-TDS experiments.

A 1-kHz regenerative laser amplifier generates 100-fs pulses, centered at 800 nm with a typical pulse energy of 3 mJ per pulse. This output is split into two replica pulses – one to generate THz by optical rectification (OR) and another to measure it by electro-optic (EO) sampling. OR occurs by weakly focusing a laser pulse into a 0.5-mm-thick, (110)-cut CSP source crystal at normal incidence with optical polarization oriented along the $[\bar{1}10]$ axis to maximize the generation [21]. A high-density polyethylene low-pass filter transmits the THz radiation, which is collected and focused onto the samples using two off-axis parabolic mirrors (OAPMs). The THz radiation transmitted through the sample is collected and refocused by two more OAPMs onto a 0.3-mm-thick, (110)-cut ZnTe EO crystal. The second laser pulse probes the THz-induced EO signal as a function of delay time ($\Delta t$), resolved by a Soleil-Babinet compensator, Wollaston polarizing prism and two balanced photodiodes (A, B) that record the THz electric field with a lock-in amplifier. The lock-in-amplifier modulation frequency is generated by mechanical chopper in the OR path and synchronized to the laser amplifier. All components from THz source to detector are enclosed in a light-tight enclosure that is purged with dry nitrogen to reduce absorption due to water vapor.

Figure 1(a) shows typical time domain results from CGP, ZGP and CSP samples at 300 K and compared to a reference signal with no sample in the cryostat. The transmitted THz pulse experiences absorption and delay due to its interaction with the sample. Time-domain results are treated with an arctan window function to remove the first reflection in ZnTe EO crystal and zero-padded the transient. Then Fourier transform is performed on the experimental transients. Figure

1 (b) shows the corresponding field amplitude spectra from a numerical Fourier transform, obtained without consideration of the large transmission delay through the samples.

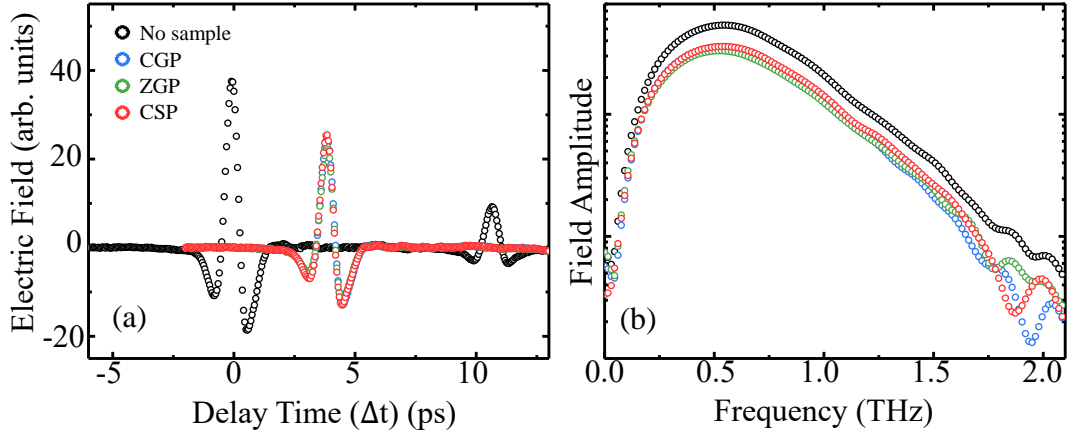

**Fig.1.** (a) Typical time-domain transient of terahertz propagating through free space (between source and detector) or transmitted through CdGeP$_2$, ZnGeP$_2$ and CdSiP$_2$ at 300 K. (b) Corresponding amplitude spectra in the frequency domain.

**Refractive Index and Absorption**

The complex transmission spectrum, $T(\omega)$, is obtained from the Fourier transform of the THz electric field through the sample, $E_{sam}(\omega)$, and free-space reference, $E_{ref}(\omega)$:

$$T(\omega) = FFT\{E_{sam}(\Delta t)/E_{ref}(\Delta t)\} = A exp(i\phi) \quad (1)$$

where $A(\omega)$ and $\phi(\omega)$ are the amplitude and spectral phase of the complex transmission. Using Fresnel equations, $T(\omega)$ can be expressed in terms of the complex refractive index, $\tilde{n}$, giving

$$T(\omega) = \frac{4\tilde{n}}{(1+\tilde{n})^2} exp[i\omega(\tilde{n}-1)d/c] \quad (2)$$

where $c$ is the speed of light in vacuum, $\omega$ is the angular frequency and $d$ is the thickness of the sample. Re-arranging equation 2, the real refractive index $n(\omega)$ and extinction coefficient $\kappa(\omega)$ are found to be [25]

$$n(\omega) = 1 + \frac{c}{\omega d}\phi(\omega) \quad (3)$$

and

$$\kappa(\omega) = \frac{-c}{\omega d} ln\left[\frac{A(1+n)^2}{4n}\right]. \quad (4)$$

In practice, to determine the correct spectral phase the large transmission delay is removed prior to performing the complex Fourier transform and additional analysis to acquire $n(\omega)$ and $\kappa(\omega)$.

Figure 2(a) shows the refractive index of all three crystals in the frequency range 0.4 -1.5 THz at 300 K. Figure 2(b) shows the THz refractive index spectra of CSP for a series of temperatures. The refractive index increases monotonically with frequency, although more significantly with temperature. The latter is due to the increased electron-phonon interactions at higher temperatures [26]. This temperature dependence can be described by the Bose-Einstein model [27,28]

$$n(T) = n_0 + S_n \left[\frac{1}{exp(\hbar\omega_{pn}/kT)-1} + \frac{1}{2}\right], \tag{5}$$

where $n_0$ is the refractive index approaching 0 K, $S_n$ is the dimensionless electron-phonon coupling constant, $k$ is the Boltzmann constant and $\omega_{pn}$ is the effective phonon frequency. The inset in Figure 2(b) shows the $n(T)$ value at 1 THz fit by the Bose-Einstein model. This models the increasing lattice vibration with increasing temperature, beyond $\hbar\omega_{pn} > kT$ near zero Kelvin. Figure 2(c) shows the resulting $S_n(\omega_{pn}/2\pi)$ revealing a linear relationship between the electron-phonon coupling with the effective phonon energy. Crystals fall on the line based on the lightest A-site cation having the highest frequency and strongest coupling, and the B-site cation influencing the frequency and coupling for the crystals with the heavier A-site-cation crystal.

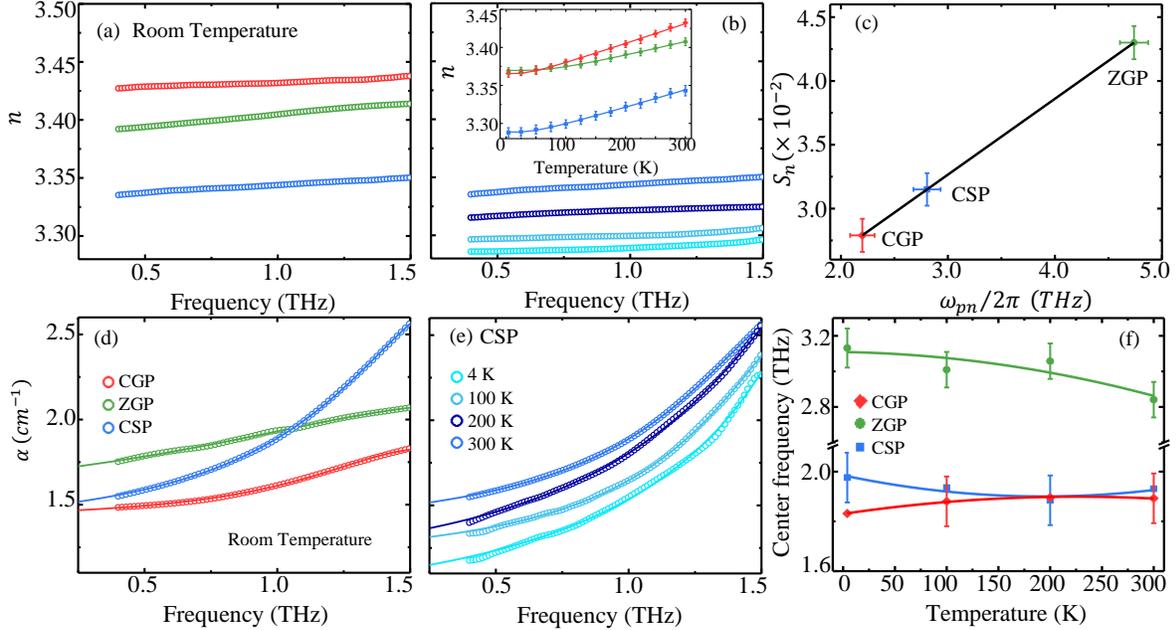

**Fig.2.** THz dispersion of refractive index for (a) CGP, ZGP, and CSP at 300 K and (b) CSP for various temperatures. The inset of (b) is the temperature-dependent refractive index at 1 THz for all three crystals, fit with a Bose-Einstein model. (c) Dependence of the extracted coupling constant versus the extracted average phonon energy with a linear fit. Absorption coefficient of (d) CGP, ZGP and CSP at 300 K and (e) CSP at various temperatures, fit with Lorentzian functions. (f) Extracted center frequency from the Lorentzian fitting as a function of temperature.

Figure 2(d) and 2(e) shows the THz absorption coefficient, $\alpha(\omega) = 2\omega\kappa(\omega)/c$, determined from the extinction coefficients. In all cases, $\alpha(\omega)$ increases with $\omega$ indicating some peak at frequencies out of the detection range. At 300 K, CSP has a more strongly varying absorption than the other crystals, which competes with a strong background absorption that is reduced by cooling. The temperature-dependent absorption results from higher free-carrier concentrations at higher temperatures [29]. A Lorentzian function is used to fit the data, with extracted center frequencies plotting in Figure 2(f). The phonon frequencies are mildly dependent on temperature with room-temperature values that match those reported from Raman measurements; namely, 1.89 THz for CGP [30], 2.82 THz for ZGP [30,31] and 1.98 THz for CSP [32].

## AC Conductivity

The complex dielectric response $\tilde{\varepsilon}(\omega)$ and complex conductivity $\tilde{\sigma}(\omega)$ are related through the complex refractive index as follows

$$\tilde{\varepsilon}(\omega) = (n + i\kappa)^2 = \varepsilon(\infty) + i\tilde{\sigma}(\omega)/\omega\varepsilon_o \tag{6}$$

where $\varepsilon(\infty)$ is the high-frequency dielectric constant and $\varepsilon_o$ is the vacuum permittivity. Rearranging equation (6) and expanding the complex conductivity to be $\tilde{\sigma}(\omega) = \sigma_1(\omega) + i\sigma_2(\omega)$ allows for determination of the real and imaginary parts. Figure 3 shows $\sigma_1(\omega)$ and $\sigma_2(\omega)$ for (a) CGP, (b) ZGP and (c) CSP both at 4 K and 300 K. For all three crystals $\sigma_1(\omega)$ is positive and increasing with frequency, whereas $\sigma_2(\omega)$ is negative and decreasing with frequency.

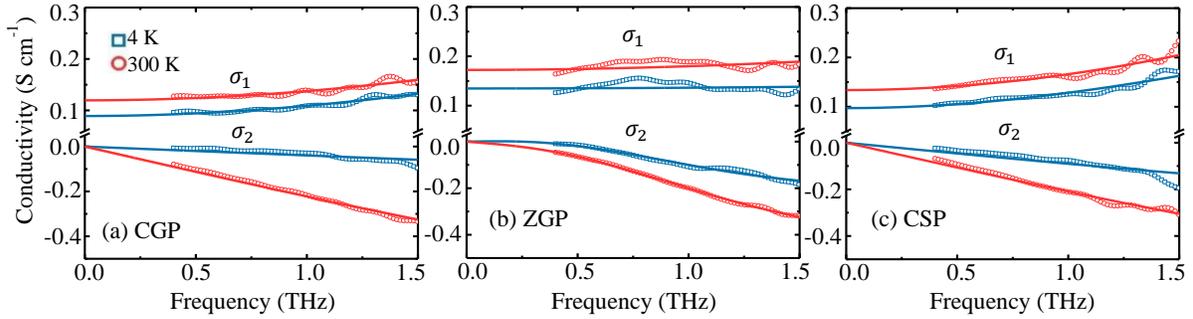

**Fig.3.** Real ($\sigma_1$) and imaginary parts ($\sigma_2$) of AC conductivity for (a) CGP (b) ZGP and (c) CSP crystals with lattice temperatures of 4 K and 300 K, fit by a Drude-Smith conduction model.

AC conductivity is related to carrier density and carrier scattering time, which in semiconductors is described by the Drude model [3]. According to this model, carrier transport is dominated by scattering events that randomize their momentum within a picosecond. Hence, $\sigma_1(\omega)$ should increase with decreasing frequency in the THz range and the $\sigma_2(\omega)$ should be negligible or even positive. The experimental trends of $\sigma_1(\omega)$ and $\sigma_2(\omega)$ are inconsistent to the pure Drude model. Smith [33] proposed a modification to the Drude model by including a carrier backscattering term associated with a negative-going photocurrent, giving

$$\sigma(\omega) = \frac{Ne^2\tau/m^\star}{1-i\omega\tau}\left[1 + \frac{c}{1-i\omega\tau}\right] \tag{7}$$

where $e$ is the elementary charge, $N$ is the carrier density, $m^\star$ is the carrier effective mass, $\tau$ is the momentum scattering time and $c$ is backscattering coefficient. A pure Drude response corresponds to $c = 0$, whereas $c \to -1$ indicates strong carrier backscattering.

The experimental AC conduction results are fit well by the Drude-Smith mode, as seen by the solid lines that overlay the data points. Figure 4 shows the fitting parameters (a) $N$, (b) $\tau$ and (c) $c$ as a function of temperature for all three crystals. In all cases, $c$ is non-negligible and reasonably independent of temperature, which indeed indicates strong carrier backscattering associated with the crystal and not carrier occupancy. In nanostructures, the carrier backscatter is easily justified by the size of the structure itself [34]. By contrast, the remaining sources of backscatter in bulk semiconductors with low polycrystallinity [24] are defects, dislocation and vacancies, which can be associated with electronic impurities (i.e., unintentional dopants) [35].

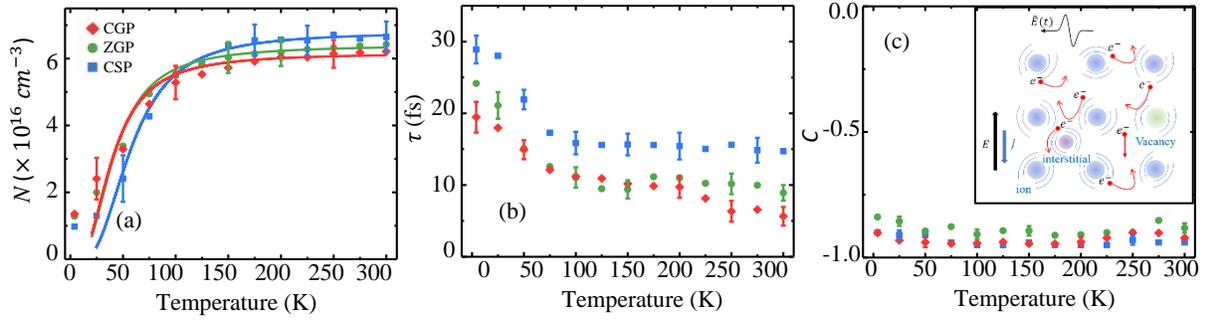

**Fig.4.** (a). Temperature-dependent carrier density (N) fit with a Fermi-Dirac model. (b)Scattering time $\tau$ as a function of temperature. (c) Backscattering coefficient for all three crystal. The inset in (c) shows a schematic diagram of the transport mechanism.

Dopants change the free-carrier occupancy as a function of temperature; hence, $N(T)$ reveals information about the scattering sources. Occupancy is described with a Fermi-Dirac model [36] including a term that aggregates all dopant species with a net dopant concentration, $N_D$, and an averaged dopant binding energy, $E_D^b$. The model is given as

$$N(T) = 2N_D / \left\{ 1 + \left[1 + 4\hat{g}(N_D/N_C)\exp(E_D^b/kT)\right]^{\frac{1}{2}} \right\} \quad (8)$$

where $N_c = 2(m_e kT/2\pi\hbar^2)^{\frac{3}{2}}$, $\hat{g} = g_1/g_0$, $g_1$ is the degeneracy of the singly-charged dopant, $g_0$ is the degeneracy of the ionized dopant, $m_e$ mass of the electron and $\hbar$ is the reduced Planck constant. $N(T)$ increases with temperature up to approximately 100 K – 150 K for the various crystals, saturating above that temperature. This trend is described by the increasing ionization of dopants up to the dopant concentration [36] and is fit well by the Fermi-Dirac model. Parameters from the model are shown in Table 1. Values of dopant concentration are reasonable and consistent with weakly-doped or compensated crystals.

Table 1. Extracted Parameters from Fermi-Dirac Model.

| Sample | $N_D(\times 10^{16} cm^{-3})$ | $E_D^B (meV)$ |
|--------|-------------------------------|---------------|
| CSP    | 6.85 ± 0.41                   | 13.13 ± 0.69  |
| ZGP    | 6.42 ± 0.32                   | 8.69 ± 0.43   |
| CGP    | 6.18 ± 0.31                   | 8.19 ± 0.41   |

The average dopant binding energy can be result from a variety of defects and vacancies and depend on the material. Electron paramagnetic measurements (EPR) of CSP crystals [37,38] have identified silicon, cadmium and phosphorus vacancies and silicon-on-cadmium anti-site defects. Silicon and cadmium vacancies acts as acceptors, whereas phosphorous vacancies and silicon-on-cadmium anti-site defects act as donors, which partly compensate one another and will exhibit different degeneracy and temperature dependencies. EPR has also observed unintentionally $Mn^+$, $Fe^+$, $Fe^{2+}$ and $Fe^{3+}$ ions in addition to the main native defects formed during growth. Silicon vacancy dominant low temperature EPR signatures and silicon-on-cadmium anti-site defects or phosphorus vacancies are more prevalent at higher temperatures. Similarly, ZGP crystals exhibit zinc, phosphorus and germanium vacancies, and germanium-on-zinc anti-site defects [39,40]. Zinc vacancy dominate the low temperature EPR response, supplanted by germanium-on-zinc anti-site or phosphorus vacancy at higher temperature. There is no information available in the literature about native defects in CGP, but they are presumed to be similar to ZGP and CSP. The temperature-dependent defect response in EPR indicated that scattering must also be temperature dependent, especially since native defects comprise donors and acceptors [37–39].

**Microscopic Scattering Mechanisms**

Carrier transform is governed by temperature-dependent scattering mechanisms that are related to either carrier or defect occupancy and/or phonon and defect interactions. Each contribution with its own temperature dependence relates the extracted scattering time and backscattering coefficient from Figure 4 (b) and (c) to the mobility and microscopic process through analysis that is typically performed on Hall measurements. Since the back-scattering coefficient is reasonably independent of temperature, the mobility can be written as $\mu(T) = e\tau(T)/m^*$ and consists of various scattering mechanism that combine by reciprocal addition, $1/\mu = \sum_i 1/\mu_i$.

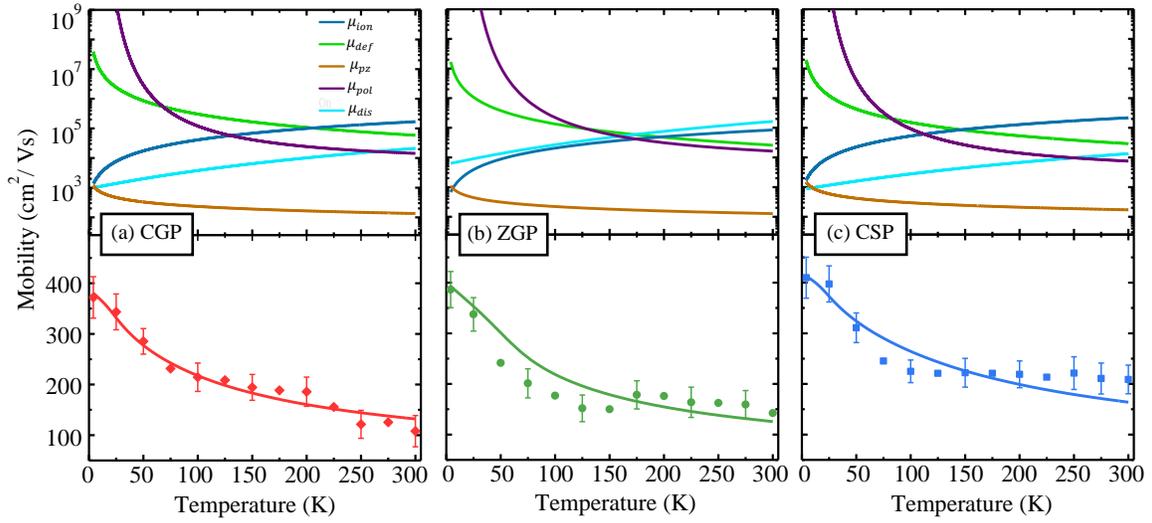

**Fig.5.** Temperature-dependent mobility fit with a multi-component scattering model for (a) CGP, (b) ZGP and (c) CSP. Solid lines represent the theoretical contributions from different scattering mechanisms.

Figure 5 shows $\mu(T)$ in the lower panels for the three crystals to decrease with increasing temperature. The data are fit with a model that incorporates several scattering mechanisms [41,42], each of which has an individual contribution that is shown in the upper panel. Ionized impurity scattering from the atomic cores is a variant of Rutherford scattering with a Coulomb potential in the Thomas-Fermi screening limit and is described as

$$\mu_{ion} = \frac{2^{7/2}(4\pi\varepsilon_0\varepsilon_{DC})^2 kT^{3/2}}{\pi^{3/2}Z^2 e^3 \sqrt{m^*} N_{ion} \ln\left[1+\left(3\varepsilon_0\varepsilon_{DC}kT/2Ze^2 N_{ion}^{1/3}\right)^2\right]} \quad (9)$$

where $\varepsilon_{DC}$ is the DC dielectric constant, $Z$ is the charge of ions and $N_{ion}$ is the ionized impurity concentration. Deformation potential scattering occurs from low wave-vector acoustic modes that depend on the elasticity of the crystal and is described as

$$\mu_{def} = \frac{2\sqrt{2\pi}e\hbar^4 c_l}{3m^{*5/2}E_{def}^2}\left(\frac{1}{kT}\right)^{3/2} \quad (10)$$

where $c_l$ is the elastic constant for longitudinal deformation and $E_{def}$ is the deformation potential. Piezoelectric potential scattering from acoustic phonons that create a piezoelectric field within the crystal is described as

$$\mu_{pz} = \frac{16\sqrt{2\pi}\hbar^2 \varepsilon_0 \varepsilon_{DC}}{3m^{*3/2} eK^2}\left(\frac{1}{kT}\right)^{\frac{1}{2}} \quad (11)$$

where $K = (e_p^2/c_l)/(\varepsilon_0 \varepsilon_{DC} + e_p^2/c_l)$ with $e_p$ as the piezoelectric coefficient. Polar optical scattering from optical phonons that have an energy comparible to $kT$ is inelastic and described as

$$\mu_{pol} = \frac{e}{2m^* \alpha \omega_0} exp\left(\frac{\theta_D}{T}\right) \quad (12)$$

where $\alpha$ is the polar constant related to the fine structure constant modified to accommodate dielectric and thermal realities of the crystal, $\omega_0$ is longitudinal optical phonon frequency and $\theta_D$ is the Debye temperature. Dislocation scattering arises from charge centers created by defect sites and is dependent on the Debye length, $\lambda_D = (\varepsilon_0 kT/e^2 N(T))^{1/2}$, and the transverse component of the kinetic energy of the carriers, $\xi_\perp = (2k_\perp \lambda_D)^2$. The mobility associates with dislocation scattering is

$$\mu_{dis} = \frac{\hbar^3 \varepsilon_0^3 c^2 eN^2}{e^3 f^2 m^{*2} N_{dis} k^2 T^2}\left(1 + \frac{4k_\perp \varepsilon_0 kT}{e^2 N}\right)^{3/2} \quad (13)$$

where $f$ is the fraction of filled traps and $N_{dis}$ the dislocation density.

Table 2. Extracted Parameters from mobility fitting for three crystals

| Parameter | CGP | ZGP | CSP |
|---|---|---|---|
| $c_l (\times 10^{11}\ Nm^{-2})$ | 1.84 | 1.12 | 1.64 |
| $E_{ac.def}$ | 8.65 | 8.10 | 7.96 |
| $e_p (Cm^{-2})$ | -3.97 | -4.85 | -5.06 |
| $\theta_D (K)$ | 319 | 391 | 376 |
| $\omega_o/2\pi (THz)$ | 6.65 | 8.15 | 7.83 |
| $\alpha$ | 0.05 | 0.03 | 0.07 |
| $N_{ion} (\times 10^{15} cm^{-3})$ | 5.36 | 9.74 | 3.57 |
| $N_{dis} (\times 10^{11}\ cm^{-2})$ | 1.87 | 0.21 | 1.26 |

The scattering model is applied to the all three crystals to show that the piezoelectric potential scattering lowers the overall mobility to current levels for nearly all temperatures and that ionized impurity contributes at low temperature. The component trends for CGP and CSP are very similar with perhaps more dislocation scattering in the former. Modelling parameters are shown in Table 2. Literature value of $\theta_D$ are used [43], $c_l$ values are in good agreement with reported values [18,44,45] and $\alpha$ and $\omega_0$ are consistent with calculations [46]. Values of $N_{ion}$ are in reasonable agreement to estimated $N_D$ from the Fermi-Dirac model presented above. Finally, to the best of the authors' knowledge, deformation potential and piezoelectric coefficient of CGP, ZGP and CSP are reported for the first time.

**Conclusion**

Temperature-dependent THz-TDS characterizes both the complex refractive index and AC conductivity and is applied to model semiconductors $CdGeP_2$, $ZnGeP_2$, and $CdSiP_2$, whose transport measurements are unreliable due to poor mobility. This approach has been used to perform spectroscopic analysis that is related to microscopic carrier transport including carrier-lattice interactions that reduce the mobility.

First, temperature-dependent THz refractive index yields the effective phonon frequency and the electron-phonon coupling constant. Second, THz absorption confirms the center frequencies of the lowest-lying optical phonon modes for each temperature. Third, temperature-dependent AC conductivity clarifies the scattering/backscattering regime, carrier concentration from dopant and their average binding energy. Finally, application of Hall-measurement analysis to the mobility (or scattering times) allows for mapping of the temperature-dependent origins of the carrier-scattering mechanisms whether that be from atoms, electrons, phonons or dislocations.

This study defines various parameters for many scattering mechanisms that are previously unreported, shows the links between non-contact THz-TDS and Hall-effect measurements. The study also supports detailed measurements of defects in these model semiconductors offering material characterization to improve bulk crystal growth for photonic and nonlinear optical applications.

**Acknowledgements**

The authors wish to thank Brett Carnio, Nancy Giles and Larry Halliburton for useful discussion.